\documentclass[12pt]{article}
\usepackage{epsfig}
\usepackage{amssymb}
\def\val{\alpha}

\def\vph{\varphi}

\newcommand{\beq}{\begin{equation}}
\newcommand{\eeq}{\end{equation}}
\newcommand{\bea}{\begin{eqnarray*}}
\newcommand{\eea}{\end{eqnarray*}}
\newcommand{\beqa}{\begin{eqnarray}}
\newcommand{\eeqa}{\end{eqnarray}}

\begin{document}
\newfont{\elevenmib}{cmmib10 scaled\magstep1}%

\newcommand{\preprint}{
           \begin{flushleft}
   \elevenmib Yukawa\, Institute\, Kyoto\\
            \end{flushleft}\vspace{-1.3cm}
            \begin{flushright}\normalsize  \sf
            YITP-02-71\\
           {\tt hep-th/0212342} \\ January 2003
            \end{flushright}}
\newcommand{\Title}[1]{{\baselineskip=26pt \begin{center}
            \Large   \bf #1 \\ \ \\ \end{center}}}
\hspace*{2.13cm}%
\hspace*{1cm}%
\newcommand{\Author}{\begin{center}\large
           P. Baseilhac\footnote{
pascal@yukawa.kyoto-u.ac.jp} \ \ and\ \ \ M.
Stanishkov\footnote{stanishkov@dante.ewha.ac.kr,
 On leave of absence from INRNE, Sofia, Bulgaria}
\end{center}}
\newcommand{\Address}{{\baselineskip=18pt \begin{center}
           \it $^{1}$Yukawa Institute for Theoretical Physics\\
     Kyoto University, Kyoto 606-8502, Japan \vspace{0.1cm}\\
     $^{2}$Department of Physics, Ewha Womans University\\
     Seoul, 120-750, Korea
     \end{center}}}
\baselineskip=14pt

\preprint
\bigskip

\Title{On the third level descendent fields in the\\ Bullough-Dodd
model and its reductions} \Author

\Address

\vskip 1.1cm

\centerline{\bf Abstract}

\vspace{0.2cm} Exact vacuum expectation values of the third level
descendent fields
 $\langle(\partial\varphi)^3({\overline\partial}\varphi)^3e^{a\varphi}\rangle$ \ in the
Bullough-Dodd model are proposed.
 By performing  quantum group restrictions, we obtain \
$\langle L_{-3}{\overline L}_{-3}{\Phi}_{lk}\rangle$ \ in perturbed minimal conformal field theories.\\

\vskip 0.8cm


\vskip -0.6cm

\vspace{-0.45cm}
\vskip 1cm

\section{Introduction}
In 2-D integrable quantum field theories which can be considered
as conformal field theories (CFTs) perturbed by a relevant
operator, two-point correlation functions are complicated objects
to study. However, using operator product expansion (OPE) in the
short-distance limit one can reduce down their expression
 in terms of vacuum expectation values (VEVs) of local fields.
Since four years, important progress has been made in this
direction, as exact VEVs either of primary fields
\cite{1,2,3,LiouI} or their first descendent \cite{des,des2,LiouI}
have been obtained explicitly. However it remains an open
important problem to find all higher level VEVs of descendent
fields and study their properties. Although a general method is
still lacking, a case by case study based on CFT data provides a
useful tool in order to determine some of the simplest higher
level VEVs. Beyond the technical aspects, the knowledge of any of
such quantities improves the analytical prediction for
short-distance expansion of two-point functions, which can be
better compared with the results obtained from the numerical study
of the model (see \cite{Num} for instance).

Recently \cite{des2}, we considered the Bullough-Dodd (BD) model
and its quantum group restrictions, following the approach of
\cite{des} concerning the sinh-Gordon or sine-Gordon models. In
Euclidean space, the action associated with the BD model writes
\beqa {\cal A}_{BD} = \int d^2x
\big[\frac{1}{16\pi}(\partial_\nu\varphi)^2 + \mu e^{b\varphi} +
\mu' e^{-\frac{b}{2}\varphi}\big]\ .\label{actionBD} \eeqa
Here, the parameters $\mu$ and $\mu'$ are introduced, as the two
operators do not renormalize in the same way, on the contrary to
any simply-laced affine Toda field theory. The purpose of this
letter is to provide an exact expression for the VEV of the third
level descendent fields (next to leading order in the UV limit of
the two-point function) in the BD model, in order to complete the
short-distance expansion of the two-point function calculated in
\cite{des2}. It should be stressed that differently from
sine-Gordon (SG) model, this VEV is nonzero, due to the existence
of a local conserved current of spin 3 in the BD model.

Finally, it is well-known that $c<1$ minimal CFT with action
\beqa {\cal{A}} &=& {\cal{M}}_{p/p'} + {\lambda} \int d^2x
\Phi_{pert}\ \label{action}\eeqa
perturbed by the operator $\Phi_{pert}\in\lbrace\Phi_{12},
\Phi_{21},\Phi_{15}\rbrace$ can be obtained by a quantum group
(QG) restriction of imaginary Bullough-Dodd model
\cite{smir,cost,tak,tak2} with special values of the coupling.
Here  we denote respectively $\Phi_{12}$, $\Phi_{21}$ and
$\Phi_{15}$ as specific primary operators of the unperturbed
minimal model ${\cal{M}}_{p/p'}$ and introduce the parameter
$\lambda$ which characterizes the strength of the perturbation.
Using this correspondence and the previous VEVs in the BD model,
we will deduce $\langle 0_s|L_{-3}{\overline
L}_{-3}\Phi_{lk}|0_s\rangle $ in the perturbed minimal model
(\ref{action}).

\section{VEVs of the third level descendent fields}
The BD  model can be regarded as a relevant perturbation of a
Gaussian CFT in which case the field is normalized such that\ \ $
\langle\varphi_(z,{\overline
z})\varphi(0,0)\rangle_{Gauss}=-2\log(z{\overline z})$. For
imaginary coupling $b=i\beta$, the perturbation is relevant for \
$0<\beta^2<1$. Although the model (\ref{actionBD}) for real
coupling is very different from the one with imaginary coupling in
its physical content (this latter model contains solitons
 and breathers), there are good
reasons to believe that the expectation values obtained in the
real coupling case provide also the expectation values for the
imaginary coupling. Then, let us now consider the two-point
function in the BD model with imaginary coupling \ \ ${\cal
G}_{\alpha_1\alpha_2}(r)=\langle
e^{i\alpha_1\vph}(x)e^{i\alpha_2\vph}(y)\rangle_{BD}$ \ with \ \
$r=|x-y|$. It can be expanded in the short-distance limit
($r\rightarrow 0$) which, as mentioned above, contains a term
corresponding to the third descendent contribution. The result
reads (see \cite{des2} for details)
\beqa &&{\cal G}_{\val_1\val_2}(r)= {\cal G}_{\val_1+\val_2}
r^{4\val_1\val_2} \Big\{ 1 + {\cal
F}_{1,2}(\val_1\beta,\val_2\beta,\beta^2)
\mu(\mu')^2r^{6-3\beta^2}  +\frac{(\val_1\val_2)^2}{4} {\cal
H}(\val_1+\val_2)r^4
\nonumber \\
&& \ \ \ \ \ \ \qquad \ \  \ \qquad \qquad \ -
\frac{\alpha_1^2\alpha_2^2(\alpha_1\!-\!\alpha_2)^2}{144}{\cal
K}(\alpha_1+\alpha_2)r^{6}
 + O\big(\mu^2(\mu')^4r^{12-6\beta^2}\big) \Big\}\nonumber\\
&& + \sum_{n=1}^{\infty} {\mu}^n
r^{4\alpha_1\alpha_2+4n\beta(\alpha_1+\alpha_2)+2n(1-\beta^2)+2n^2\beta^2}
j_n(\val_1\beta,\val_2\beta,\beta^2){\cal
G}_{\val_1+\val_2+n\beta} \ \Big\{1+
O\big(\mu(\mu')^2r^{6-3\beta^2}\big)\Big\}\nonumber
 \\
&& + \sum_{n=1}^{\infty} {\mu'}^n
r^{4\alpha_1\alpha_2-2n\beta(\alpha_1+\alpha_2)+2n(1-\frac{\beta^2}{4})+\frac{n^2\beta^2}{2}}
j_n(-\frac{\val_1\beta}{2},-\frac{\val_2\beta}{2},\frac{\beta^2}{4}){\cal
G}_{\val_1+\val_2-\frac{n\beta}{2}}
\ \Big\{1+ O\big(\mu(\mu')^2r^{6-3\beta^2}\big) \Big\}\nonumber \\
&& + \sum_{n=1}^{\infty} {\mu}^n{\mu'}
r^{4\alpha_1\alpha_2+4(n-\frac{1}{2})\beta(\alpha_1+\alpha_2)+2n(1-2\beta^2)+2+2n^2\beta^2}
{\cal F}_{n,1}(\val_1\beta,\val_2\beta,\beta^2)\nonumber{\cal
G}_{\val_1+\val_2+(n-\frac{1}{2})\beta}\nonumber\\
 && \qquad \ \qquad \qquad \qquad \qquad \qquad \qquad \qquad \qquad \qquad \qquad \qquad \times \ \Big\{1+
O\big(\mu(\mu')^2r^{6-3\beta^2}\big)\Big\}\label{twop} \eeqa
where we defined ${\cal H}(\alpha)$ and ${\cal K}(\alpha)$ by the
ratios
\beqa {\cal H}(\alpha)= \frac{\langle
(\partial\vph)^2({\overline\partial}\vph)^2e^{i\alpha\vph}\rangle
_{BD}} {\langle e^{i\alpha\vph}\rangle _{BD}}\ \label{H}\qquad
{\mbox{and}}\ \qquad {\cal K}(\alpha)= \frac{\langle
(\partial\vph)^3({\overline\partial}\vph)^3e^{i\alpha\vph}\rangle
_{BD}} {\langle e^{i\alpha\vph}\rangle _{BD}}\label{Ka} \eeqa
and\ ${\cal G}_\val=\langle e^{i\alpha\vph}\rangle _{BD}$ is the
VEV of the exponential field in the BD model. A closed analytic
expression for  ${\cal G}_\val$ and ${\cal H}(\alpha)$ has been
proposed in ref. \cite{2} and ref. \cite{des2}, respectively.
Their expression involves an integral representation which is well
defined if
\beqa -\frac{1}{2\beta}\ <\ {\mathfrak R}e(\alpha)\ <\
\frac{1}{\beta}\label{cond} \eeqa
and obtained by analytic continuation outside this domain. Here we
used the notations of \cite{des2} for the Dotsenko-Fateev
integrals $j_{n}(a,b,\rho)$ and ${\cal F}_{n,m}(a,b,\rho)$. In
particular, the integrals $j_n(a,b,\rho)$ have been evaluated
explicitly in \cite{dots}  with the result
\beqa j_{n}(a,b,\rho)=\pi^n\prod_{k=0}^{n-1}
\frac{\gamma((k+1)\rho)}{\gamma(\rho)}\gamma(1+2a+k\rho)
\gamma(1+2b+k\rho) \gamma(-1-2a-2b-(n-1+k)\rho)\label{jn} \eeqa
where the notation $\gamma(x)=\Gamma(x)/\Gamma(1-x)$ is used.
Also, the integral ${\cal F}_{1,1}(a,b,\rho)$ can be obtained from
the result of \cite{mag}. Instead, the integral ${\cal
F}_{1,2}(a,b,\rho)$ is a quite complicated object, and its
explicit calculation goes beyond the purposes of this letter.

In the (Gaussian) free field theory, the composite fields
$(\partial\vph)^3({\overline\partial}\vph)^3e^{i\alpha\vph}$ are
spinless with scale dimension
\beqa D\equiv \Delta+{\overline \Delta}=2\alpha^2+6\ .\label{dim}
\eeqa
For $0<\beta^2<1$ the perturbation is relevant and a finite number
of lower scale dimension counterterms are sufficient to cancel the
divergences arising in the VEVs of third level descendent fields.
However, this procedure is regularization scheme dependent, i.e.
one can always add finite counterterms. For generic values of
$\alpha$ this ambiguity in the definition of the renormalized
expression for these fields can be eliminated by fixing their
scale dimensions to be (\ref{dim}). It exists however a set of
values of $\alpha$ for which the ambiguity still remains. In the
BD model with imaginary coupling, this situation arises if two
fields, say ${\cal O}_{\val}$ and ${\cal O}_{\val'}$, satisfy the
resonance condition
\beqa D_\val=D_{\val'} + 2n(1-\beta^2) + 2n'(1-\frac{\beta^2}{4})\
\ \ \ \mbox{with}\ \ \ \ (n,n')\in {\mathbb N} \eeqa
associated with the ambiguity
\beqa {\cal O}_{\val}\longrightarrow{\cal O}_{\val} +
{\mu}^{n}{\mu'}^{n'}{\cal O}_{\val'}\ . \eeqa
In this case one says that the renormalized field ${\cal O}_\val$
has an $(n|n')$-th {\it resonance} \cite{des,des2} with the field
${\cal O}_{\val'}$. Due to the condition (\ref{cond}) we find
immediately that a resonance can appear between the third level
descendent field
$(\partial\varphi)^3(\bar\partial\varphi)^3e^{i\alpha\varphi}$ and
the following primary fields:
\beqa (i) \ \ \ \ &&e^{i(\val-\beta)\vph}\ \ \ \ \mbox{i.e.}\ \
(n|n')=(1|4) \ \ \
\mbox{for}\ \ \ \alpha=\frac{1}{\beta}-\frac{\beta}{2}\ ;\label{nm}\\
(ii) \ \ \ \ &&e^{i(\val+3\beta)\vph}\ \ \ \mbox{i.e.}\ \
(n|n')=(3|0) \ \ \
\mbox{for}\ \ \ \alpha=-\beta\ ;\nonumber\\
(iii) \ \ \ \ &&e^{i(\val-\frac{\beta}{2})\vph}\ \ \ \
\mbox{i.e.}\ \ (n|n')=(0|1) \ \ \
\mbox{for}\ \ \ \ \alpha=-\frac{2}{\beta}\ ;\nonumber\\
(iv) \ \ \ \ &&e^{i(\val-\frac{3\beta}{2})\vph}\ \ \ \
\mbox{i.e.}\ \  (n|n')=(0|3) \ \ \ \mbox{for}\ \ \
\alpha=\frac{\beta}{2}\ .\nonumber \eeqa
If we now look at the expression (\ref{twop}), we notice that the
contribution brought by the third level descendent field in
(\ref{Ka}), and that of any of the exponential fields in  $(i)$,
$(ii)$, $(iii)$ and $(iv)$, have the same power behavior in $r$
($r^{4\val_1\val_2 + 6}$) at short-distance for the corresponding
values of $\alpha$. The integrals which appear in these
contributions are, respectively:
 \beqa &&(i)\ \qquad {\cal
F}_{1,4}(\val_1\beta,\val_2\beta,\beta^2)\ ,\qquad \
\ \  (ii)\ \ \ j_3(\val_1\beta,\val_2\beta,\beta^2)\ ,\nonumber\\
&&(iii)\ \ \
j_1(-\frac{\val_1\beta}{2},-\frac{\val_2\beta}{2},\frac{\beta^2}{4})\
,\qquad \ (iv)\ \
j_3(-\frac{\val_1\beta}{2},-\frac{\val_2\beta}{2},\frac{\beta^2}{4})\
. \nonumber\eeqa

As we will see, ${\cal K}(\val)$ (and similarly for the real
coupling case) exhibits the same poles in order that the divergent
contributions compensate each other. This last requirement leads
for instance to a set of relations for ${\cal K}(\alpha)$. The
third one reads
\beqa \frac{\val_1^2\val_2^2(\val_1-\val_2)^2}{144}{\cal
R}es_{\alpha=-\frac{2}{\beta}}\ {\cal K}(\val) = \mu'\frac{{\cal
G}_{\val-\beta/2}}{{\cal G}_\val}|_{\val=-\frac{2}{\beta}}\ {\cal
R}es_{\alpha=-\frac{2}{\beta}} \
j_1(-\frac{\val_1\beta}{2},-\frac{\val_2\beta}{2},\frac{\beta^2}{4})\
,\label{residu} \eeqa
which is used to fix the $\val$-independent part (normalization)
of ${\cal K}(\val)$.

On the other hand, to determine the explicit form of the
$\val$-dependent part of ${\cal K}(\val)$, we use the reflection
relations method. Indeed, the BD model (\ref{actionBD}) can be
regarded as two different perturbations of the Liouville field
theory \cite{2}. First, one can consider the Liouville action
where the perturbation is identified with $e^{-\frac{b}{2}\vph}$.
The holomorphic stress-energy tensor
\beqa T(z)=-\frac{1}{4}(\partial\vph)^2 +
\frac{Q}{2}\partial^2\vph\label{str} \eeqa
which ensures the local conformal invariance of the Liouville
field theory with coupling $b$ can be written in terms of the
standard Virasoro generators\ \ $T(z)=\sum_{n\in{\mathbb Z}}L_n
z^{-n-2} \ \
 \mbox{and} \  \  {\overline T}({\overline z})=\sum_{n\in{\mathbb Z}}{\overline L}_n
{\overline z}^{-n-2}$.
Then, using the OPE of the stress-energy tensor of the Liouville
part with any primary field, we have the relation
\beqa
 L_{-3}{\overline L}_{-3}e^{a\varphi}=\big[\big(\frac{a+Q}{2}\big)^2\partial^3\varphi
-\frac{1}{2}\partial^2\varphi\partial\varphi\big]\big[\big(\frac{a+Q}{2}\big)^2{\overline\partial}^3\varphi
-\frac{1}{2}{\overline\partial}^2\varphi{\overline\partial}\varphi\big]e^{a\varphi}\
. \eeqa
Furthermore, taking the expectation value of the combination above
and using the (Gaussian) equations of motion $\partial{\overline
\partial} \vph=0$ we obtain
\beqa \langle L_{-3}{\overline L}_{-3}e^{a\varphi}\rangle
_{BD}=\frac{a^2}{16}(a+1/b)^2(a+b)^2
\langle(\partial\varphi)^3({\overline\partial}\varphi)^3e^{a\varphi}\rangle
_{BD}\ . \label{L3L3} \eeqa
Alternatively, we can consider $e^{b\varphi}$ as a perturbation.
Using both pictures and CPT framework, we deduce reflection
relations between operators with the same quantum numbers. We
report the reader to \cite{2,3,des,des2} for details about this
approach. Consequently, if we denote
\beqa
K(a)=\frac{\langle(\partial\varphi)^3({\overline\partial}\varphi)^3e^{a\varphi}\rangle
_{BD}} {\langle e^{a\varphi }\rangle_{BD}}\ ,\label{Kreal} \eeqa
then we obtain the following two functional
relations
\beqa
K(a)&=&\Big[\frac{(b+1/b-a)(b+2/b-a)(2b+1/b-a)}{a(a+1/b)(a+b)}\Big]^2K(Q-a)\ ,\label{ref}\\
K(-a)&=&\Big[\frac{(b/2+2/b-a)(b/2+4/b-a)(b+2/b-a)}{a(a+2/b)(a+b/2)}\Big]^2K(-Q'+a)\
.\nonumber \eeqa
Notice that these equations are invariant with respect to the
symmetry
 $b\rightarrow-\frac{2}{b}$ with $a\rightarrow -a$ in agreement with the
well-known self-duality of the BD-model. Assuming that $K(a)$ is a
meromorphic function in $a$, we find that the ``minimal'' solution
which follows from (\ref{residu}), (\ref{ref}) is:
\beqa
K(a)\!&=&\!-\frac{1}{a^2}\Big[\frac{m\Gamma(\frac{b^2}{h})\Gamma(\frac{2}{h})}{\Gamma(\frac{1}{3}){\sqrt
3}\ 2(Q+Q')^2}
\Big]^6\gamma\big(\frac{2ba+b^2+2}{h}\big)\gamma\big(\frac{-2ba-2}{h}\big)
\gamma\big(\frac{2ba-b^2+4}{h}\big)\gamma\big(\frac{-2ba-2b^2}{h}\big)\nonumber\\
\nonumber &&  \ \ \ \qquad \qquad \qquad \qquad \quad\  \times \ \
\gamma\big(\frac{-2ba+2b^2-2}{h}\big)\gamma\big(\frac{2ba-4}{h}\big)
\gamma\big(\frac{-2ba+b^2+2}{h}\big)\gamma\big(\frac{2ba-b^2}{h}\big)\label{L3L3}
\eeqa
where $h=6+3b^2$ is the ``deformed'' Coxeter number
\cite{DGZ,CDS}. Here we have used the exact relation between the
parameters $\mu$ and $\mu'$ in the action (\ref{actionBD}) and the
mass of the fundamental particle $m$ \cite{2} :
\beqa m=\frac{2\sqrt 3 \Gamma(1/3)}{\Gamma(1+b^2/h)\Gamma(2/h)}
\big( -\mu\pi\gamma(1+b^2)\big)^{1/h} \big(
-2\mu'\pi\gamma(1+b^2/4)\big)^{2/h}.\label{massmu} \eeqa
Notice that $K(a)$ is invariant under the duality transformation
${b\rightarrow -2/b}$ as expected, and contains all the expected
poles. Accepting this conjecture and taking $a=0$, we obtain for
instance:
\beqa\langle L_{-3}{\overline L}_{-3}{\mathbb
I}\rangle_{BD}=-\frac{m^2}{2^{10/3}}\frac{\Gamma^2(1+2/h)\Gamma^2(1+b^2/h)\Gamma^2(2/3)}
{\gamma(1/2+2/h)\gamma(1/2+b^2/h)\gamma(1/3+6/h)\gamma(1/3+3b^2/h)}f_{BD}^2\eeqa
where $f_{BD}$ is the bulk free energy of the Bullough-Dodd model,
obtained in \cite{2}.
\section{Application to perturbed conformal field theories}
For imaginary value of the coupling $b=i\beta$, with the
substitutions $\mu\rightarrow-\mu$ and $\mu'\rightarrow -\mu'$ in
(\ref{actionBD}) the BD model possesses quantum group symmetry
$U_q(A_2^{(2)})$ with deformation parameter $q=e^{i\pi/\beta^2}$
\cite{smir,cost}. At roots of unity, it is used to describe
$\Phi_{12}$, $\Phi_{21}$ or $\Phi_{15}$ perturbed CFTs
(\ref{action}). Let us consider
 the first case i.e. the $\Phi_{pert}\equiv\Phi_{12}$ perturbation,
  obtained for $\beta^2=p/p'$ \ with \ $ 1<p<p'$ relative prime
integers. In the following, $\Phi_{lk}$ will denote a primary
field of the minimal model ${\cal M}_{p/p'}$. The exact relation
between the parameters $\lambda$ in (\ref{action}) and the mass of
the fundamental kink $M$ can be found in \cite{2}. Here we denote
\beqa \xi=\frac{p}{p'-p}\ .\label{xi} \eeqa
For unitary minimal models $\xi>1$ which, for ${\mathfrak
I}m(\lambda)=0$, corresponds to a massive phase \cite{2}. Using
the particle-breather identification \cite{2}\
$m=2M\sin\big(\frac{\pi\xi}{3\xi+6}\big)$ \ and parameter
$a=i\big(\frac{l-1}{2\beta}-\frac{k-1}{2}\beta\big)$ in $K(a)$ it
is then straightforward to get the VEV:
\beqa \frac{\langle0_s|L_{-3}{\overline
L}_{-3}\Phi_{lk}|0_s\rangle}{\langle0_s|\Phi_{lk}|0_s\rangle} &=&
-\Big[\frac{2^{2/3}\pi M\Gamma(\frac{2+2\xi}{3\xi+6})}{{\sqrt
3}\Gamma(\frac{1}{3})\Gamma(\frac{\xi}{3\xi+6})(1+\xi)}\Big]^6
\frac{1}{\xi^2(1+\xi)^2(3\xi+6)^2}\nonumber\\
&&\ \ \ \times\ \ \frac{\gamma(\frac{\eta-4\xi-3}{3\xi+6})
\gamma(\frac{-\eta-4\xi-3}{3\xi+6})\gamma(\frac{\eta+1+\xi}{3\xi+6})\gamma(\frac{-\eta+1+\xi}{3\xi+6})
}{\gamma(\frac{\eta+2\xi+3}{3\xi+6})
\gamma(\frac{-\eta+2\xi+3}{3\xi+6})\gamma(\frac{\eta-2\xi+1}{3\xi+6})\gamma(\frac{-\eta-2\xi+1}{3\xi+6})}
\ .\label{12} \eeqa
Here $|0_s\rangle$ is one of the degenerate ground states of the
QFT (\ref{action}) (see \cite{2} for a detailed discussion of the
vacuum structure of the model).

For the second restriction $\beta^2=p'/p$ which leads to the
action (\ref{action}) with $\Phi_{pert}\equiv\Phi_{21}$, the exact
relation between the parameter $\lambda$ and the mass of the
fundamental kink $M$ has been obtained in \cite{2}. The VEV of the
third order descendent field immediately follows from (\ref{12})
with the replacement $\xi\rightarrow -1-\xi$.

Another subalgebra of $U_q(A^{(2)}_2)$ is the subalgebra
$U_{q^4}(sl_2)$. One can again restrict the phase space of the
complex BD with respect to this subalgebra for a special value of
the coupling $\beta^2=4p/p'$ \ with \ $2p<p'$ relative prime
integers in order to describe the third case, i.e.
$\Phi_{pert}=\Phi_{15}$. The calculations are straightforward so
we will not report them here.

\section{Conclusion}
In conclusion, we have conjectured an exact expression for the VEV
of the third level descendent in the BD model. The highly
nontrivial check of the residue conditions corresponding to the
poles (\ref{nm}), (\ref{residu}) strongly supports our conjecture.

As explained above, the computation of the (UV behavior) of the
two-point function involves an infinite tower of VEVs of
descendent fields. It is not clear how to solve this problem in
general. Even in the simplest case as the SG theory, a system of
functional equations appear for the 4-th level descendent. A
solution of this problem is still an open question.

Yet, for practical reasons, the computation of any higher order
descendent VEV gives new information. Our results for example can
be used in improving the comparison with the numerical
computations in some interesting statistical models around their
critical point: the critical Ising model in a magnetic field
\cite{Num}, the tricritical Ising model perturbed by its energy
operator \cite{mag} or by its subleading magnetic operator,...

\paragraph*{Aknowledgements}
We are grateful to Al. B. Zamolodchikov for discussions. MS's work
is supported by KISTEP exchange program 12-69-002. PB's work is
supported by JSPS fellowship.


\begin{thebibliography}{10}
%

\bibitem{1} S. Lukyanov and A. B. Zamolodchikov, Nucl. Phys.  {\bf B 493} (1997) 571.
%
\bibitem{2}
V. A. Fateev, S. Lukyanov, A. B. Zamolodchikov and Al. B.
Zamolodchikov, Nucl. Phys. {\bf B 516} (1998) 652.
%
\bibitem{3}
P. Baseilhac and V. A. Fateev, Nucl. Phys. {\bf B 532} (1998)
567;\\
V. A. Fateev, Mod. Phys. Lett. {\bf A 15} (2000) 259;\\
C. Ahn, P. Baseilhac, V. A. Fateev, C. Kim and C. Rim, Phys. Lett.
{\bf B 481} (2000) 114;\\
C. Ahn, P. Baseilhac, C. Kim and C. Rim, Phys. Rev. {\bf D 64}
(2001) 046002.
%
\bibitem{LiouI}
P. Baseilhac, Nucl. Phys. {\bf B 636} (2002) 465.
%
\bibitem{des}
V. A. Fateev, D. Fradkin, S. Lukyanov, A. B. Zamolodchikov and Al.
B. Zamolodchikov, Nucl. Phys. {\bf B 540}, (1999) 587.
%
\bibitem{des2}
P. Baseilhac and M. Stanishkov, Nucl.Phys. {\bf B 612} (2001) 373.
%
\bibitem{Num}
M. Caselle, P. Grinza and N. Magnoli, Nucl. Phys. {\bf B 579}
(2000) 635;\\
M. Caselle, P. Grinza and N. Magnoli, J. Phys. {\bf A 34} (2001)
8733.
%
\bibitem{smir} F. Smirnov, Int. J. Mod. Phys. {\bf A 6} (1991) 1407.
%
\bibitem{cost} C. J. Efthimiou, Nucl. Phys. {\bf B 398} (1993) 697.
%
\bibitem{tak}
M. J. Martins, Phys. Lett. {\bf B 262} (1991) 39;\\
A. Koubek, M. J. Martins and G. Mussardo, Nucl. Phys. {\bf B 368}
(1992) 591.
%
\bibitem{tak2}
G. Takacs, Nucl. Phys. {\bf B 489} (1997) 532;\\
H. G. Kausch, G. Takacs and G. M. T. Watts, Nucl. Phys. {\bf B
489} (1997)
557;\\
G. Takacs and G. M. T. Watts, Nucl. Phys. {\bf B 547} (1999) 538.
%
\bibitem{dots}
V. S. Dotsenko and V. A. Fateev, Nucl. Phys. {\bf B 240} (1984) 312;\\
V. S. Dotsenko and V. A. Fateev, Nucl. Phys. {\bf B 251} (1985)
691.
%
%
\bibitem{mag}
R. Guida and N. Magnoli, Int. J. Mod. Phys. {\bf A 13} (1998)
1145.
%
\bibitem{DGZ}
G. W. Delius, M. T. Grisaru and D. Zanon, Nucl. Phys. {\bf B 382}
(1992) 365.
%
\bibitem{CDS}
E. Corrigan, P. E. Dorey and R. Sasaki, Nucl. Phys. {\bf B 408}
(1993) 579.
%

\end{thebibliography}
\end{document}